\documentclass[12pt,preprint]{aastex}
\usepackage{natbib}
\usepackage{amsmath}
\usepackage{mathrsfs}
\newcommand{\dvcs}{$\delta V_{CS}$}
\newcommand{\hill}{\sc hill5}
\newcommand{\thco}{$^{13}$CO}
\newcommand{\ceo}{C$^{18}$O}
\newcommand{\nht}{NH$_3$}
\newcommand{\cts}{C$_2$S}

\newcommand{\nthp}{N$_2$H$^+$}

\newcommand{\kms}{km\,s$^{-1}$}
\newcommand{\cmt}{cm$^{-3}$}
\newcommand{\msun}{$M_{\odot}$}

\newcommand{\core}{L1551~MC}
\slugcomment{Accepted for Publication in the Astrophysical Journal}
\shorttitle{A Pre-Protostellar Core in L1551. II}
\shortauthors{Swift et al.}
\begin{document}
\title{A Pre-Protostellar Core in L1551. II. State of Dynamical and
  Chemical Evolution}
\author{Jonathan J. Swift\altaffilmark{1} and William J. Welch}
\affil{Department of Astronomy and Radio Astronomy Laboratory,
University of California, 601 Campbell Hall, Berkeley, CA
94720-3411}

\author{James Di Francesco}
\affil{National Research Council of Canada, Herzberg Institute of
  Astrophysics, 5071 West Saanich Road, Victoria, BC V9E 2E7, Canada}

\author{Irena Stojimirovi\'c}
\affil{Department of Astronomy, University of Massachusetts, LGRT-B
  619E, 710 North Pleasant St., Amherst, MA 01003-9305}

\altaffiltext{1}{\bf {\tt js@astro.berkeley.edu}}
\begin{abstract}
Both analytic and numerical radiative transfer models applied to
high spectral resolution CS and {\nthp} data give insight into the
evolutionary state of {\core}. This recently discovered
pre-protostellar core in L1551 appears to be in the early stages of
dynamical evolution. Line-of-sight infall velocities of $\gtrsim
0.1$\,{\kms} are needed in the outer regions of {\core} to adequately
fit the data. This translates to an accretion rate of $\sim
10^{-6}$\,{\msun}\,yr$^{-1}$, uncertain to within a factor of 5 owing
to unknown geometry. The observed dynamics are not due to spherically
symmetric gravitational collapse and are not consistent with the
standard model of low-mass star formation. The widespread, fairly
uniform CS line asymmetries are more consistent with planar
infall. There is modest evidence for chemical depletion in the radial
profiles of CS and {\ceo} suggesting that {\core} is also chemically
young. The models are not very sensitive to chemical
evolution. {\core} lies within a quiescent region of L1551 and is
evidence for continued star formation in this evolved cloud.
\end{abstract}

\keywords{ISM: clouds --- ISM: individual(\objectname{L1551}) ---
  stars: formation --- line: profiles --- radiative transfer}

\section{Introduction}
Gravitationally bound molecular cores with no embedded infrared
sources, referred to as pre-protostellar cores (PPCs), offer the
unique opportunity to study the conditions preceding stellar birth. To
date there have only been a few PPCs studied in depth leaving many 
unanswered questions regarding the kinematics and chemistry during
this epoch of star formation. 

Inward gas motions are necessary for the formation of stars and the
kinematic signatures of these motions are expected to be present in
molecular line data. Blue-peaked asymmetric line profiles commonly
seen in optically thick tracers of molecular cores can be accurately
modeled as self-absorbed infalling gas~\citep{eva99,mye96}. A
high ratio of blue-peaked to red-peaked profiles in selected cores is
further evidence that these spectral signatures are tracing global
infall in some PPCs~\citep{lee01}. These ``infall asymmetries'' imply
infall velocities of up to $\sim 0.1$\,{\kms} extending out to $\sim
0.1$\,pc~\citep{taf98,lee01} inconsistent with the ambipolar diffusion
paradigm of isolated low-mass star
formation~\citep{shu87,cio95}.

The observed decrement in abundance of certain molecules such as CO
and CS toward the centers of cores~\citep{kui96,wil98,cas99,taf02} can
be explained by the expected depletion of these species onto dust
grains~\citep{wat72,ber97}. In some PPCs the abundance profile is best
fit using a model with a central region of complete
depletion~\citep[e.g.,][]{taf04}. Models have shown that depletion
becomes more severe as a core collapses suggesting that chemical
depletion may be useful in determining the age of
PPCs~\citep{raw92,ber97,lee04,aik05}. However, detailed molecular
line studies have shown that chemistry is not reliable as a sole
evolutionary indicator~\citep{lee03} and interpretations should be
made on a case by case basis~\citep{raw01}.

Analytic models of asymmetric molecular line
profiles~\citep[e.g.][]{mye96} provide quick solutions by imposing
functional forms for molecular excitation and chemistry along the line
of sight. These models have been shown to reliably characterize
simulated data despite using the simplest available
assumptions~\citep{dev05}. Numerical models typically contain less
assumptions and are hence more flexible, but are computationally much
more costly.  

{\core} is a PPC that sits in a relatively isolated region to the
northwest of IRS5 in the well-known L1551 dark 
cloud~\citep[][hereafter Paper~I]{swi05}. It is expected to be
gravitationally unstable and hints of infall have been seen in
position switched CS spectra. The differentiation seen between the
{\nht} and {\cts} emission suggests chemical evolution. Past studies
of the L1551 cloud~\citep[e.g.,][]{sne81,dev99} provide a
well-defined context for {\core}. Following the discovery and primary
characterization of this starless molecular core (Paper~I), this
article investigates dynamical and chemical signatures from new
molecular line data in an attempt to understand its state of evolution
toward stellar birth.

In \S\,\ref{sObs} we outline the molecular line observations used with
the observations of Paper~I to analyze {\core}. The models used to
interpret these data are described in \S\,\ref{sModels} with the
results presented in \S\,\ref{sResults}. Further consideration of
these results in \S\,\ref{sDisc} reveal the main conclusions of this
paper which are then summarized in \S\,\ref{sConc}. 

\section{Observations and Reductions} \label{sObs}
We used the Five College Radio Astronomy Observatory's 14\,m telescope
on the nights of 2005 January 7 and 8, and for partial nights on
February 8, 9 and 24 to map {\core} in \nthp and CS simultaneously.
The weather was variable on the nights of January 7 and 8 with system
temperatures varying from $\sim 220$\,K to $\sim 375$\,K. The partial
nights were substantially better showing system temperatures from
$130$\,K to $150$\,K.

The observing frequencies in the first and second IF of the dual
channel correlator (DCC) were set to 93.176258\,GHz and
97.980953\,GHz, corresponding to the isolated component of the
{\nthp}$(1-0)$ and the CS$(2-1)$ rotational transitions~\citep{lee01}
respectively, for our maps. The beam at these observing frequencies
has a FWHM of $\sim 45${\arcsec} and a main beam efficiency $\eta_b
\approx 0.5$. The highest spectral resolution mode of the DCC was used
giving a 25\,MHz band with 1024 channels resulting in $\sim
0.08$\,{\kms} velocity resolution.

A $6\arcmin \times 6\arcmin$ patch of sky was mapped centered on
{\core}, $4^{\rm h}31^{\rm m}09^{\rm s}\!.9$,
$+18^\circ12^\prime41^{\prime\prime}\,$(J2000.0), using the on-the-fly
observing technique. Our final maps are a composite of $\sim20$
individual maps created by scanning alternatively in the right
ascension and declination directions combined by weighted mean. These
maps are highly oversampled given the 12{\arcsec} row spacing,
60{\arcsec}\,s$^{-1}$ scan speed and the rotation of the SEQUOIA 32
element receiver array with paralactic angle. The data are
calibrated with an {\sc off} scan every two rows and a total power
calibration scan every four rows. These data were reduced, gridded and
output to FITS files using the {\sc otftool} utility available at the
observatory. The RMS noise level in our final CS and {\nthp} maps is
$\sim 26$\,mK. All further analysis was performed using the IDL
software of Research Systems Inc.

\section{Models} \label{sModels}
We use a non-linear least squares minimization based on the
Levenberg-Marquardt technique to find the best fits to our spectral
line data according to two distinctly different radiative transfer
models. The analytic model fits the asymmetric CS line profiles while
the family of numerical models use information from the optically thin
{\nthp} emission to fit both the CS and the CO isotopologue data.

\subsection{Analytical Model}
The {\hill} model presented in \cite{dev05} assumes a symmetric
1-dimensional structure for all parameters. The central excitation
temperature, $T_C$, tapers linearly to the cloud edge where it has a
value of $T_b = 2.73$\,K. Gas approaches the center of this model
system at a constant velocity, $v_{in}$, relative to the systemic
velocity, $v_{LSR}$. Including the central optical depth, $\tau_C$,
and the velocity dispersion of the gas, $\sigma$, there are a total of
5 free parameters. The {\hill} model is the preferred model in
~\cite{dev05} giving estimates of the infall velocity in simulated
observations good to 0.02\,{\kms}.

\subsection{Numerical Models} \label{sNmodels}
The ``{\sc rt}'' program introduced in \cite{dic94} \citep[also
see][]{wil04} computes the full, non-equilibrium radiative transfer
for a spherically symmetric mass. This {\sc fortran} code takes the
density, kinematics, kinetic temperature and molecular abundance ratio
as a function of core radius and solves for the level populations
using a lambda-iteration technique.

Table~\ref{da94table} outlines the fixed and free parameters which
define our family of numerical models and Table~\ref{da94models}
summarizes the individual models described in the following
subsections. The density of molecular hydrogen is represented by 
$n$. The density of a molecular species is denoted by $n_{mol}$ and
is related to the total density by the conversion factor,
$X_{mol}$. The infall velocity, $v_{in}$, is equal to $v_0$ at the
minimum radius of the model and follows a power law form with index
$p$. The model core is isothermal with a constant turbulent velocity
width $\Delta v_{turb}$. Quantities are evaluated at 40
logarithmically spaced radii from $r_{min}$ to $r_{max}$.  The fixed
parameters, except for the radii, are derived from the {\nht} data of
Paper~I. The maximum radius represents the extent of the CS
emission. The minimum radius was chosen in consideration of a timely
convergence and our results are not sensitive to its exact value. All
models have as free parameters $n_0$, $v_0$, and $X_{mol}$.   

Each numerical model comprises a substantial parameter space to 
explore. The density and infall velocities are therefore
constrained to $10^3 < n_0 < 10^7$\,{\cmt} and $0 < v_{in} <
0.5$\,{\kms} in accord with the results of Paper~I. For
each model the best fit was determined by repeating the least-squares
minimization procedure beginning at randomly chosen parameter values
until the global minimum in $\chi^2$ space was converged upon. 

These models were first used to fit the CS$(2-1)$ line profiles and
then modified to fit for the conversion factors for {\thco} and {\ceo}
given the best fit parameters from the CS fits. The kinetic
temperature and turbulent velocity width are also free parameters in
these fits to account for the possibility that the CO isotopologues are
tracing different environments within {\core}. Upper limits of
12\,K~\citep{sne81} and 0.5\,{\kms} were enforced for $T_K$ and
$\Delta v_{turb}$ respectively.

The systemic velocity for all the numerical models is obtained
from fits to the full {\nthp}$(1-0)$ transition with the hyperfine
fitting routine developed in Paper~I. This routine returns the optical 
depth, velocity centroid, velocity width, and excitation temperature
for a given spectrum. 

The normalized root mean square of the error is used as our goodness of
fit indicator defined by
\begin{equation}
RMSE = \frac{1}{\sigma}\sqrt{\frac{1}{N}\sum_{i=1}^{N}(y_i-\hat{y}_i)^2}
\label{rmse}
\end{equation}
where $\sigma$ is the RMS noise level in the spectrum, $N$ is the
number of data points used in the fit, $y_i$ are the data points and
$\hat{y}_i$ are the model points. Values of $RMSE \sim 1$ are
expected for successful fits. 

There are six total numerical models denoted by an identifier and a
Roman numeral (Table~\ref{da94models}). Each
pair of models are identical except for the abundance profile which is
either constant throughout or takes the form
\begin{equation}
X_{mol} = 
\begin{cases} 
   0             & : \quad r_{min} \le r \le r_c \\
  {\rm constant} & : \quad r_c < r \le r_{max} \\
\end{cases}
\label{drop}
\end{equation}
which is referred to as a ``drop'' abundance profile. The models are
further grouped by the assumed velocity structure into two main
categories, accretion and collapse.

\subsubsection{``Accretion'' Models}
The first four models are referred to as ``accretion'' models because
they allow for the possibility of high velocity gas in the outer
regions of the core. Model {\sc acv-i} is like the {\hill}
analytic model, but the radiative transfer is solved for
numerically. Therefore, the infall velocity from {\sc acv-i} can be
compared to the results of the {\hill} model for consistency. Model
{\sc acv-ii} is used to test whether or not having a depletion zone
significantly alters the derived infall velocity. 

Models {\sc a-i} and {\sc ii} have a velocity profile that increases
as the square root of the core radius and represent scenarios in which
gas is accreting onto the core from large radii. These models, in
comparison with the others, test if the highest velocity infalling gas
is in the exterior regions of the core. 

\subsubsection{``Collapse'' Models}
Models {\sc gc-i} and {\sc ii} have velocity structures chosen to
represent inside-out collapse, as in \cite{shu77}. Given the
complexity of {\core} from observations a detailed comparison with
theoretical models would be difficult. However, free-fall collapse is
generally expected to produce a decreasing velocity profile with
radius from the center of the core to the rarefaction
front~\citep[e.g.,][]{fat04,mye05}. Therefore these models test if the
motions implied by the CS emission are consistent with the motions
expected from gravitational collapse directly associated with the
formation of a protostar. 

\section{Results} \label{sResults}
\begin{figure}
\centering
\includegraphics[width=6in]{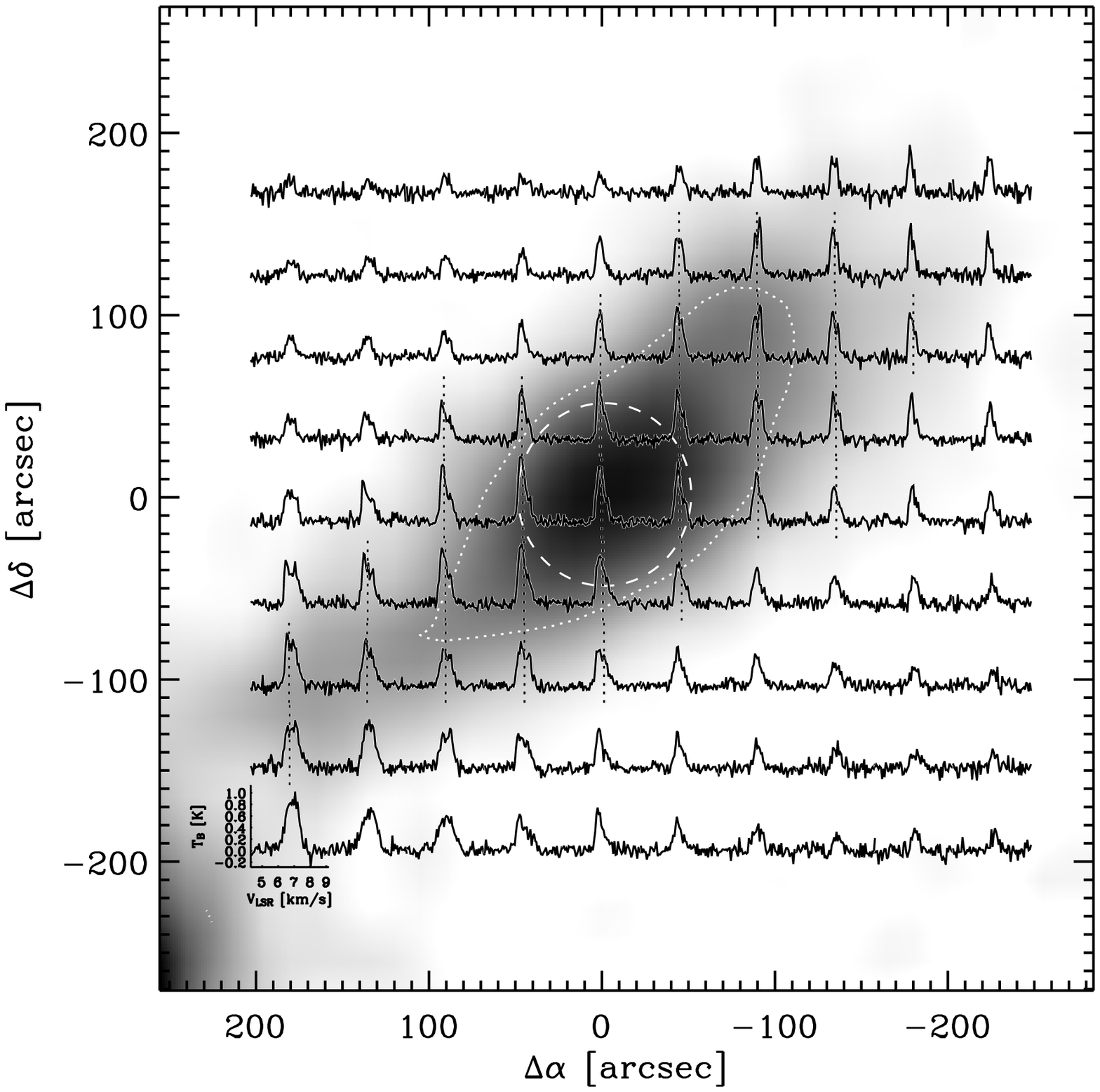}
\caption{Velocity integrated {\nthp} emission from {\core} in
greyscale with independent spectra of CS$(2-1)$ overlaid at their
respective positions. The dotted contour represents the 50\% peak
{\nthp} emission contour. The dashed circle denotes the region in
which composite spectra were constructed for {\nthp}, CS, {\thco}, and
{\ceo}. The vertical dotted lines in selected spectra represent the
systemic velocity along the line-of-sight determined from fits to the
{\nthp} spectra at the respective locations.  \label{specoverlay}}
\end{figure}

Figure~\ref{specoverlay} shows the distribution of velocity integrated
{\nthp} in greyscale overlaid with independent CS$(2-1)$ spectra at
their respective plane-of-sky locations. The vertical dotted lines
shown in selected spectra are the systemic velocities from the
hyperfine fits to the {\nthp} profiles. Most CS line profiles are
asymmetric having a peak blueward of the systemic velocity and a
smaller red peak or shoulder.

\subsection{Statistics of the CS Line Asymmetry} \label{sStats}
The normalized velocity difference introduced in \cite{mar97} is a
useful quantity to characterize spectral asymmetries. Using CS and
{\nthp} as optically thick and thin molecular species, we define
\begin{equation}
\delta V_{CS} = \frac{V_{CS} - V_{N_2H^+}}{\Delta V_{N_2H^+}}
\label{dvcs}
\end{equation}
where $V_{CS}$ is the velocity centroid of the brightest peak in the
CS line profile and the velocity centroid and intrinsic FWHM of the
{\nthp} emission, $V_{N_2H^+}$ and $\Delta V_{N_2H^+}$, are obtained
from hyperfine fitting.

\begin{figure}
\centering
\includegraphics[angle=90,width=6in]{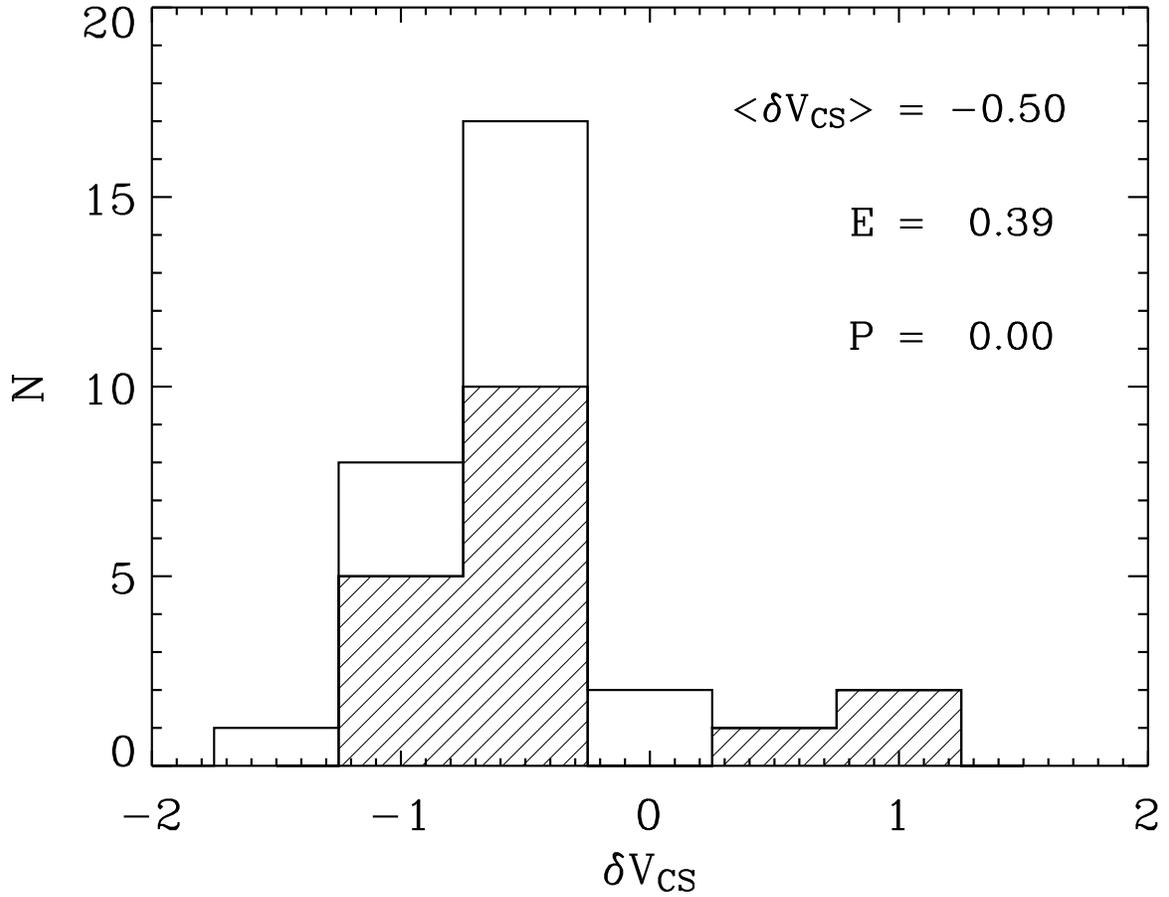}
\caption{Histogram of {\dvcs} values for all spectra with reliable
velocity centroids determined from fits to the {\nthp} data. The
shaded region corresponds to the histogram of spectra with
$|\delta V_{CS}| \ge 5\sigma_\delta$. See \S\,\ref{sStats} for
details. \label{dvcshist}}
\end{figure}

The {\nthp} emission is less extended than the CS emission, so it is
the 31 beam-sampled spectra with well-determined centroids
($\sigma_{V_{cen}} < 0.1\Delta V_{N_2H^+}$) that define our
sample. Figure~\ref{dvcshist} shows a histogram of the
{\dvcs} values. The cross-hatched area corresponds to {\dvcs} values
that are greater than 5 times the estimated error ($\sigma_\delta$)
obtained from standard propagation of error. The values for
$\sigma_\delta$ range from 0.03 to 0.4 with a median of 0.10. The mean
value $\langle${\dvcs}$\rangle = -0.50$.

The blue excess is defined as $E = (N_{-}-N_{+})/N$~\citep{lee01},
where $N_{-}$ is the number of positions with {\dvcs} $\le
-5\sigma_\delta$, $N_{+}$ is the number of positions with {\dvcs} $\ge
5\sigma_\delta$ and here $N = 31$. The blue excess of {\core} is high,
$E = 0.39$, meaning a significant fraction of CS lines have a blue
peak.

The $P$ value of the student t-test measures of the probability that a
given sample of values was drawn from a normal distribution with a
mean of zero. For the 31 {\dvcs} values in {\core}, $P$ is less
than 0.5\%. Together with a high $E$ value, this means that the
asymmetric CS profiles are not random but are strongly biased toward
having a blue peak. In relation to the cores studied in \cite{lee01},
{\core} would rank as a bona fide infall candidate. 

\subsection{Model Fits} \label{sModelFits}
The 18 spectra with $|${\dvcs}$| \ge 5\sigma_\delta$ are fit according
to the {\hill} model. Three of these 18 spectra return positive
(outflowing) gas velocities and correspond precisely to the three
spectra that show positive {\dvcs} values in Figure~\ref{dvcshist}. Of
the 15 spectra that show negative (infalling) 
gas velocities, the mean $\langle v_{in}\rangle = 0.13$\,{\kms},
roughly two thirds the sound speed in the medium.

\begin{figure}
\centering
\includegraphics[height=6in]{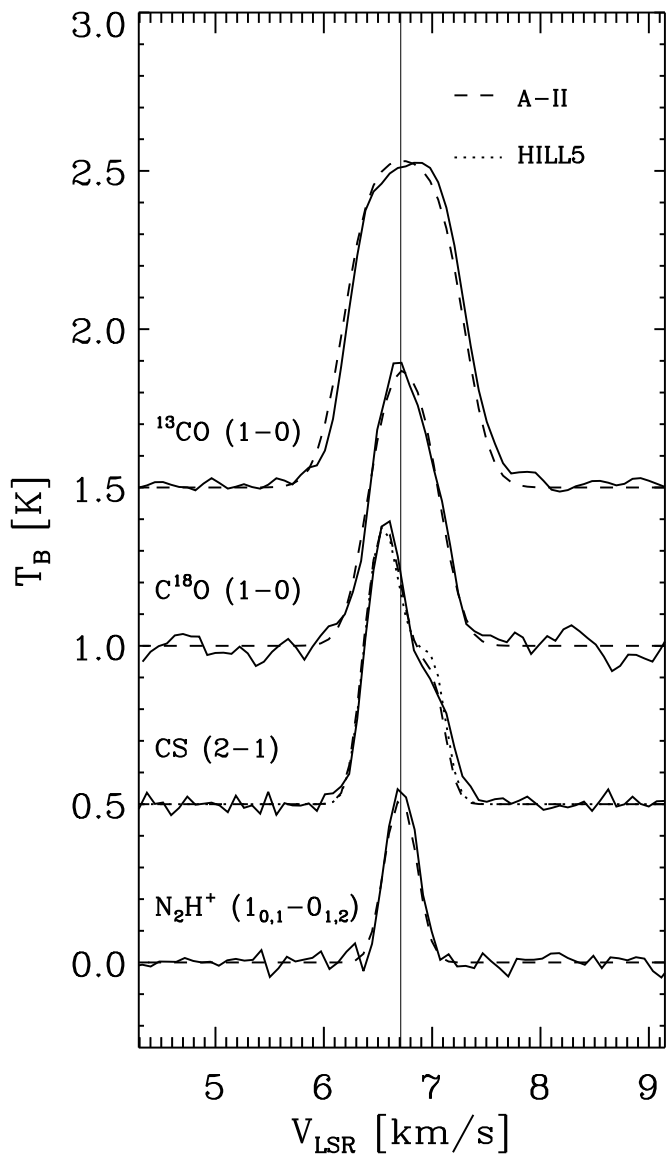}
\caption{{\nthp}, CS, {\ceo}, and {\thco} composite spectra in
  brightness temperature units. The scale of the {\ceo} and {\thco}
  spectra are reduced by a factor of 2 and 5 respectively. The model
  {\sc a-ii} fits to all spectra and 
  {\hill} fit to the CS spectrum are shown as dashed and dotted lines
  respectively. The thin vertical line signifies the velocity centroid
  of the {\nthp} line. \label{specfits}} 
\end{figure}

Averages of our molecular line data within the central 100{\arcsec} of
{\core} (dashed circle in Figure~\ref{specoverlay}) yield high
signal-to-noise composite spectra in CS, {\nthp}, {\thco}, and
{\ceo}. The best fit to the CS composite spectrum using the {\hill}
model is outlined in Table~\ref{hillfittable} and is presented as the
dotted line on the CS spectrum of Figure~\ref{specfits}. The fit is
good overall, but slightly overestimates the emission in the red wing
of the profile.

The results from the numerical fits are summarized in
Table~\ref{daresults}. The best fit to the CS composite spectrum was
achieved with model {\sc a-ii} with an RMSE value of 0.77 and is shown
as the dashed line on the CS spectrum of Figure~\ref{specfits}. The
constant velocity models, {\sc acv-i} and {\sc ii} fit the data fairly
well and are in agreement with the {\hill} results, while the collapse
models, {\sc gc-i} and {\sc ii}, produce the worst agreement with the
data.

\section{Discussion} \label{sDisc}
\subsection{Dynamics}
There is a surplus of blue peaked asymmetric CS line profiles in
{\core} successfully modeled as self-absorption from globally
infalling gas (\S\,\ref{sResults}). In both our analytic and
numerical models, gas velocities $\gtrsim 0.1$\,{\kms} are needed 
to account for the observed CS profiles. These motions are not
associated with any of the known outflows in the L1551 cloud,
including those from IRS5~\citep{sne80,mor88,mor91}, L1551
NE~\citep{pou91,mor95,dev99} and XZ/HL Tau~\citep{wel00}. A closer
look at our results reveals insights into the nature of these observed
dynamics.

\subsubsection{``Collapse'' vs. ``Accretion''} \label{sCvsA}
\begin{figure}
\centering
\includegraphics[angle=90,width=6in]{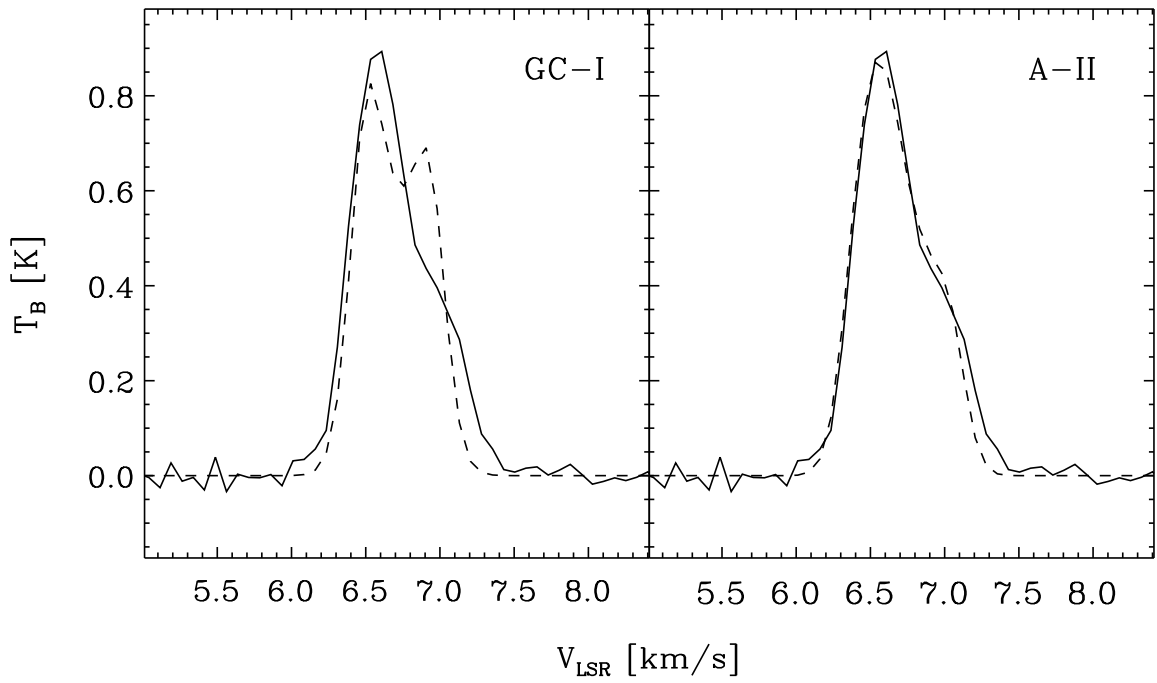}
\caption{Comparison between the best fit ``collapse'' model, 
{\sc gc-i}, to the best fit ``accretion'' model, {\sc a-ii}. The solid
line denotes the data while the model fits are represented with the
broken line. See \S\,\ref{sCvsA} for details. \label{compare}} 
\end{figure}

Figure~\ref{compare} shows a direct comparison between the best
``collapse'' model fit, {\sc gc-i}, and the best ``accretion'' model
fit, {\sc a-ii}. The {\sc gc} models produce the least consistent fits
to our data. In these models the absorbing gas is
in the outer regions of the core where the excitation temperatures are
lowest and where gas is moving at relatively slow
velocities. Therefore, these models produce centralized
self-absorption with a double peaked profile instead of the observed
red shoulder. Higher values of $v_0$ push the absorption feature
slightly further to the red and this is why the best fits for the
gravitational collapse models are pegged against the $v_0 =
0.5$\,{\kms} limit set by our constraint on highly supersonic gas
(\S\,\ref{sNmodels}).

The constant velocity accretion models, {\sc acv-i} and {\sc ii}, do
substantially better than the inside-out collapse models allowing for
absorption and emission of high velocity gas. But the monotonically
increasing infall velocity of the {\sc a-i} and {\sc ii} accretion
models works together with the decreasing CS excitation temperature as
a function of radius to carve a smooth red shoulder in the line
profile most consistent with our data. A marginally better fit to the
data is achieved with the drop abundance profile of {\sc a-ii} and we
use this as the preferred model.

\subsubsection{High Velocity Gas at Large Core Radii}
The observed self-absorption in the CS line profiles forces our
numerical model fits into the optically thick regime. It is
therefore the dynamics, rather than the chemistry or mass
distribution, that most significantly affects the emergent line
profiles. Our fits show that infalling gas is needed in the outer
regions of the core to reproduce the observed CS line profiles. The
{\hill} model fits to the ensemble of CS spectra also show high infall
velocities at large projected core radii (see \S\,\ref{sAccretion}).

According to the standard model of low-mass star
formation~\citep{shu87}, a core collapses via ambipolar diffusion
until it becomes supercritical and then undergoes an inside-out
collapse. The high infall velocities at large core radii indicated by
our models are comparable to other closely studied cores where the
dynamics were found to be inconsistent with the standard
model~\citep[e.g.][]{taf98,lee01}. From our models it is clear that
the gas motions observed across {\core} do not represent infalling gas
due to gravitational collapse and have magnitudes too large to be
associated with ambipolar diffusion.

The high infall velocities observed over a large angular extent around
{\core} are more consistent with planar accretion. Planar accretion would
produce this observational signature for a wide range of inclination
angles. The magnetic field around L1551 has been measured to be fairly
uniform~\citep{vrb76} and a roughly constant field could facilitate a
planar accretion scenario. However, measurements of the magnetic field
geometry in {\core} are essential to clarify this situation.  

From the analytic and numerical modeling of {\nthp} and CS
spectra in {\core} it appears that gas is falling toward the mid-plane
of the core from large radii. The observed dynamics are not directly
related to the formation of a star, but do not preclude gravitational
collapse scenarios~\citep[e.g.,][]{fat04}.

\subsubsection{Mass Accretion} \label{sAccretion}
High infall velocities are derived from the {\hill} model for a vast
majority of the CS spectra which span $\sim$6\arcmin, or 0.24\,pc at
the distance of Taurus. This loosely defines a radius at which material
is infalling, $R_{in} \sim 0.1$\,pc.

The mass infall rate is estimated as the amount of material passing
through a spherical surface with radius $R_{in}$, $\dot{M}_{in} = 4
\pi R_{in}^2 \rho v_{in}$. Here we use a density of
$1.2\times10^4$\,{\cmt} and an infall velocity of 0.02\,{\kms} derived
from the model {\sc a-ii} best fit parameters and the density and
velocity relations from Table~\ref{da94table} evaluated at
$R_{in}$. The derived infall rate $\dot{M}_{in} \sim
10^{-6}$\,{\msun}\,yr$^{-1}$. Values of the average density
(Paper~I) and average infall velocity (\S\,\ref{sModelFits}) yield an
accretion rate of the same order of magnitude as does planar infall at
a constant velocity of 0.1\,{\kms}. Due to the unknown
geometry and symmetry in this problem the accretion rate cannot be
constrained within a factor of $\sim 5$.

\subsection{Chemical Structure}
\subsubsection{Comparisons of Molecular Species}
A fit to the mean spectrum within a half power contour of the {\nthp}
velocity integrated emission gives a central velocity of
$6.71$\,{\kms}, an observed line width of $0.28$\,{\kms} translating
into a non-thermal width of $0.25$\,{\kms} for a kinetic temperature
of 9\,K and a total optical depth of 4.03 ($\tau = 0.4$ for the
isolated component). This is very similar to the {\nht} emission with
$v_{LSR} = 6.72$\,{\kms} and $\Delta v_{nt} =
0.2$\,{\kms}~(Paper~I). The sky distribution of {\nthp} also
follows the {\nht} closely with a major and minor FWHM of
$2.\!^\prime9$ and $1.\!^\prime3$ oriented at a position angle of
$139^\circ$ compared to the {\nht} values of $2.\!^\prime3$,
$1.\!^\prime1$ and $133^\circ$. The emission from both molecular
species peak at the same sky coordinates and coincide precisely with
a peak in the sub-mm dust continuum~\citep{sch04}. The
{\nthp} velocity gradient across the core is 1.3\,{\kms}\,pc$^{-1}$ at
a position angle of $227^\circ$ compared to the {\nht} values of
1.2\,{\kms}\,pc$^{-1}$ and $224^\circ$. Therefore it seems the {\nthp}
and {\nht} emission trace the same high-density interior gas in {\core}. 

\begin{figure}
\centering
\includegraphics[angle=90,width=6in]{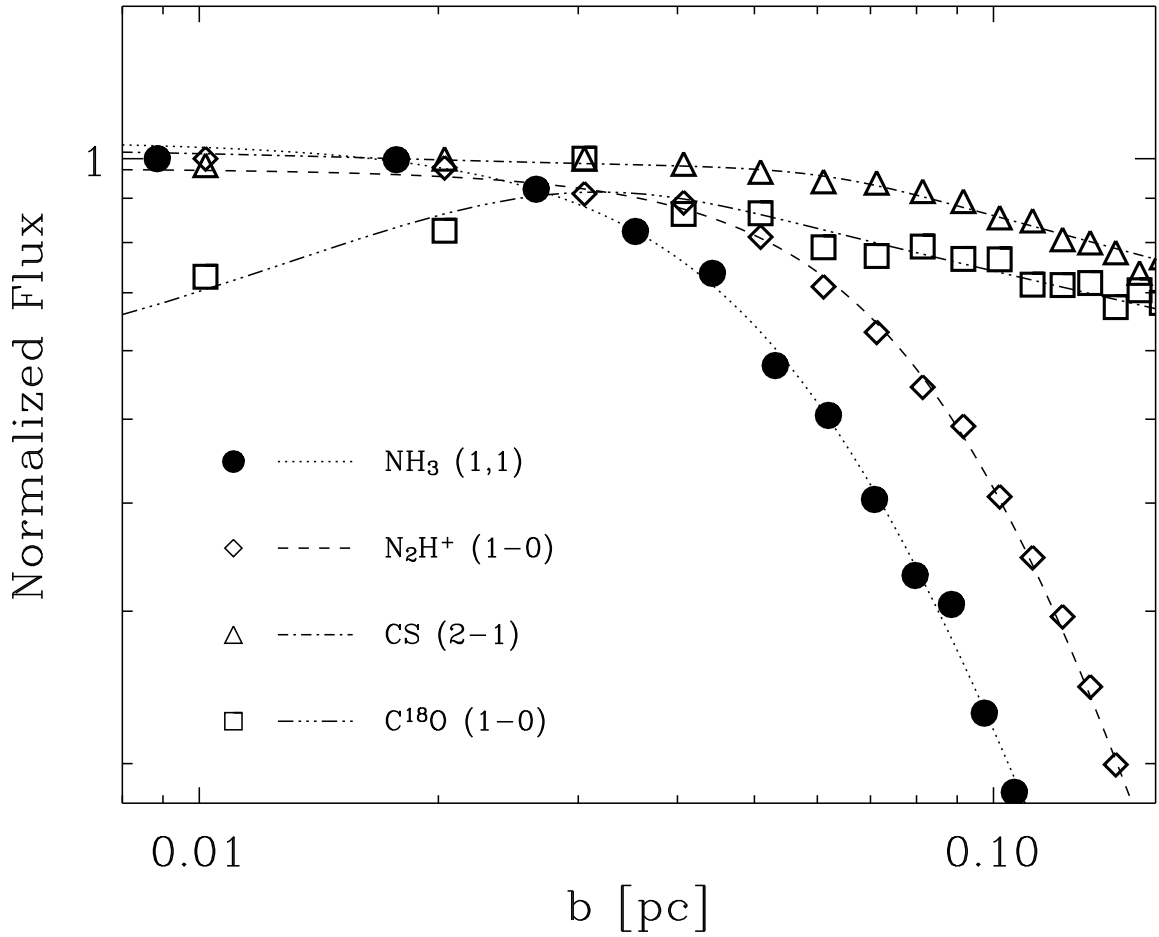}
\caption{Profiles of four different molecular species averaged in
  elliptical annuli around {\core}. For each species the flux is
  normalized to the peak value in the profile. Equation 1 (for
  {\nht} and {\nthp}) and Equation 2 (for CS and {\ceo}) from
  Paper~I are used to fit the profiles. \label{profiles}} 
\end{figure}

\begin{figure}
\centering
\includegraphics[angle=270,width=6in]{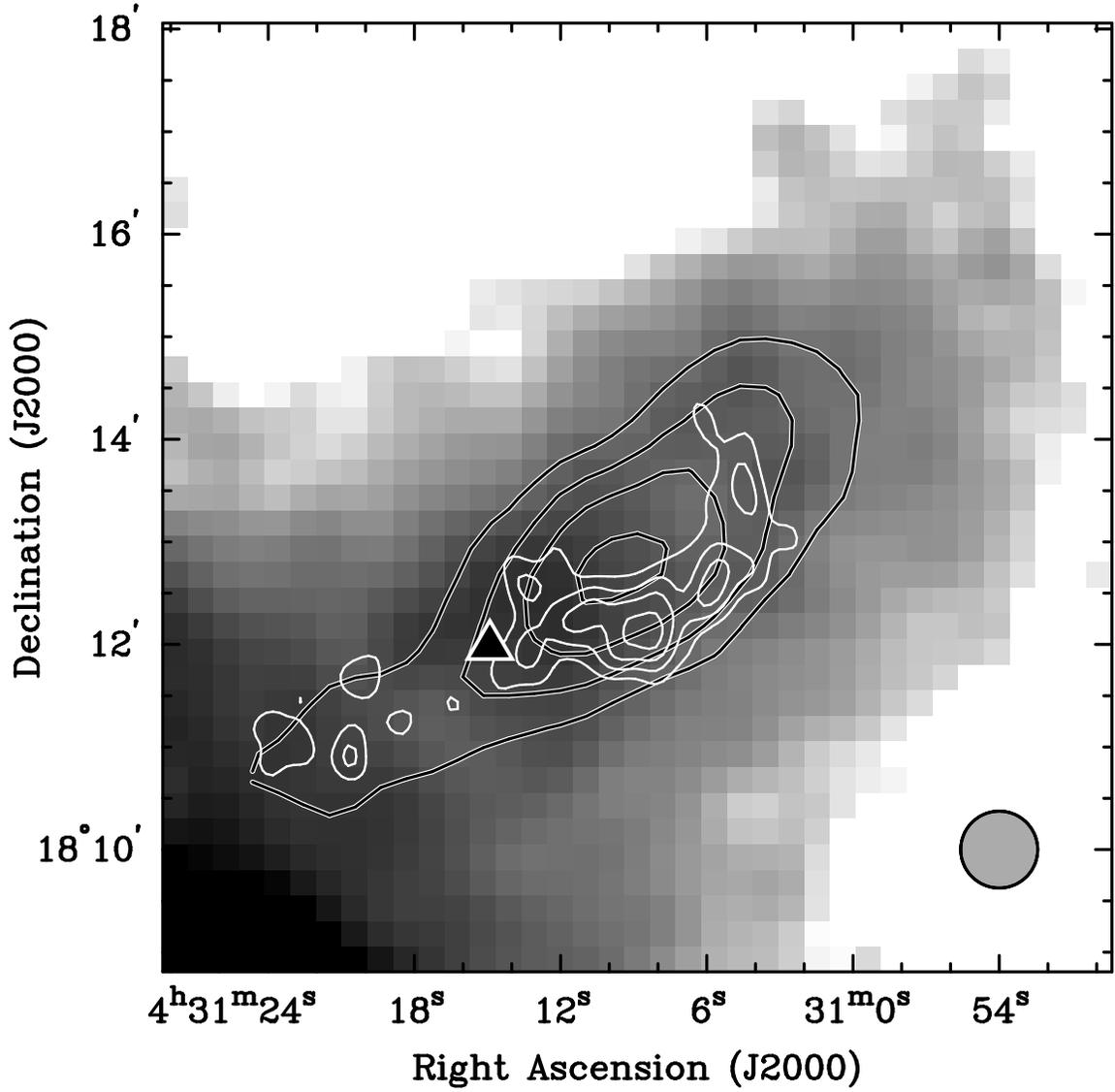}
\caption{Velocity integrated CS in greyscale overlaid with contours of
{\nthp} (black contours) and {\cts} (white contours; Paper I) at 30,
50 70 and 90\% peak flux. The beam size for the CS and {\nthp} data is
shown at the bottom right. The black triangle denotes the position of
HH265. \label{velintcs}}
\end{figure}

Figure~\ref{profiles} compares the radial profiles of {\nht}, {\nthp},
CS, and {\ceo}. The {\nht} and {\nthp} profiles are fit using Equation
1 and the CS and {\ceo} profiles are fit using Equation 2 of
Paper~I. The radial distribution of {\nthp} is slightly flatter than
the {\nht} which may be due to a slight overabundance of {\nht} in the
innermost regions as seen in other Taurus
cores~\citep[e.g.][]{taf04}. The CS and {\ceo} profiles are
significantly flatter than the {\nht} and {\nthp} suggesting that the
relative abundance of these species is higher at larger radii. While
this may be due in part to the optical depth of CS, the {\ceo} is
optically thin in this region (see \S\,\ref{sCO}). Figure~\ref{velintcs}
shows that the CS has a much broader distribution on the sky compared
to the {\nthp} despite the similarity in critical density values for
the transitions. A depression in the CS emission $\sim 1\arcmin$
northwest of the peak {\nthp} and {\nht} emission is also seen in
Figure~\ref{velintcs} that coincides with a deficit in {\ceo}
emission. This slight depression is bordered on the southwestern edge
by the {\cts} emission and may signify a depletion zone.

It is not clear whether or not significant depletion has occurred in
the inner regions of {\core} since chemical effects cannot be
sufficiently isolated with our models. However, the relative 
deficit of emission from both {\ceo} and CS at small radii
indicate some depletion of these species has occurred presumably due to
chemical evolution related to core contraction~\citep[e.g.][]{lee04}.

\subsubsection{CO Isotopologue Emission} \label{sCO}
Figure~\ref{specfits} shows the best fit to the composite CO
isotopologue spectra and Table~\ref{dacotable} outlines the fit
results. The best fit turbulent velocities for our CO isotopologues
are a factor of 1.5 higher than the CS and {\nthp} fits. These larger
values are likely due to the CO isotopologues tracing a larger volume
along the line of sight. The CO isotopologues are tracing cold gas
with the {\thco} and {\ceo} being optically thick and thin
respectively.

The abundance of both isotopologues derived from our models are
consistent with observations~\citep[e.g.,][]{fre82,lad94} given the
stated uncertainties. The best fits for $X_{^{13}{\rm CO}}$ and
$X_{{\rm C}^{18}{\rm O}}$ are $1.1 \times 10^{-6}$ and $1.1 \times
10^{-7}$ respectively. For all our models the ratio of these
abundances most consistent with the observed CO isotopologue line
intensities is $\approx 10$; a factor of 2 higher than what would be
estimated from terrestrial abundance ratios~\citep{pic98}. 

The centroid of the {\ceo} line agrees well with the centroid of the
{\nthp} (see Figure~\ref{specfits}) in accord with the results from
\cite{lee01}. The shape of the {\thco} line profile, slightly
broadened by optical depth, is fit well with model {\sc a-ii}, but the
central velocity is clearly shifted. This shift accounts for the
anomalously high RMSE values in Table~\ref{dacotable}. A cross
correlation between the model {\sc a-ii} fit and the observed spectrum
peaks at a lag of 7.1\,kHz corresponding to a velocity offset of
0.02\,{\kms} at this transition frequency. This is not accounted for
in the stated measurement errors for the published frequency equal to
$5.1$\,kHz~\citep{pic98}. If this offset were due to an error in the
chosen observing frequency, a best estimate of the true frequency
would be 110.2013470\,GHz. 

\subsection{Is the {\cts} Emission Tracing the CS Self-Absorbing Gas?}
It is likely that CS and {\cts} share the same chemical
pathway in the interstellar medium~\citep[e.g.,][]{suz92}. Yet the
critical density of CS$(2-1)$ is an order of magnitude higher than
{\cts}$(3_2 \rightarrow 2_1)$~\citep{wol97}. It is therefore possible
for these species to exist in the same environment in different
excitation states. The {\cts} emission in {\core} is shifted to the
red of the systemic velocity (see Paper~I) while the CS spectrum shows
redshifted self-absorption. 

\begin{figure}
\centering
\includegraphics[angle=90,width=6in]{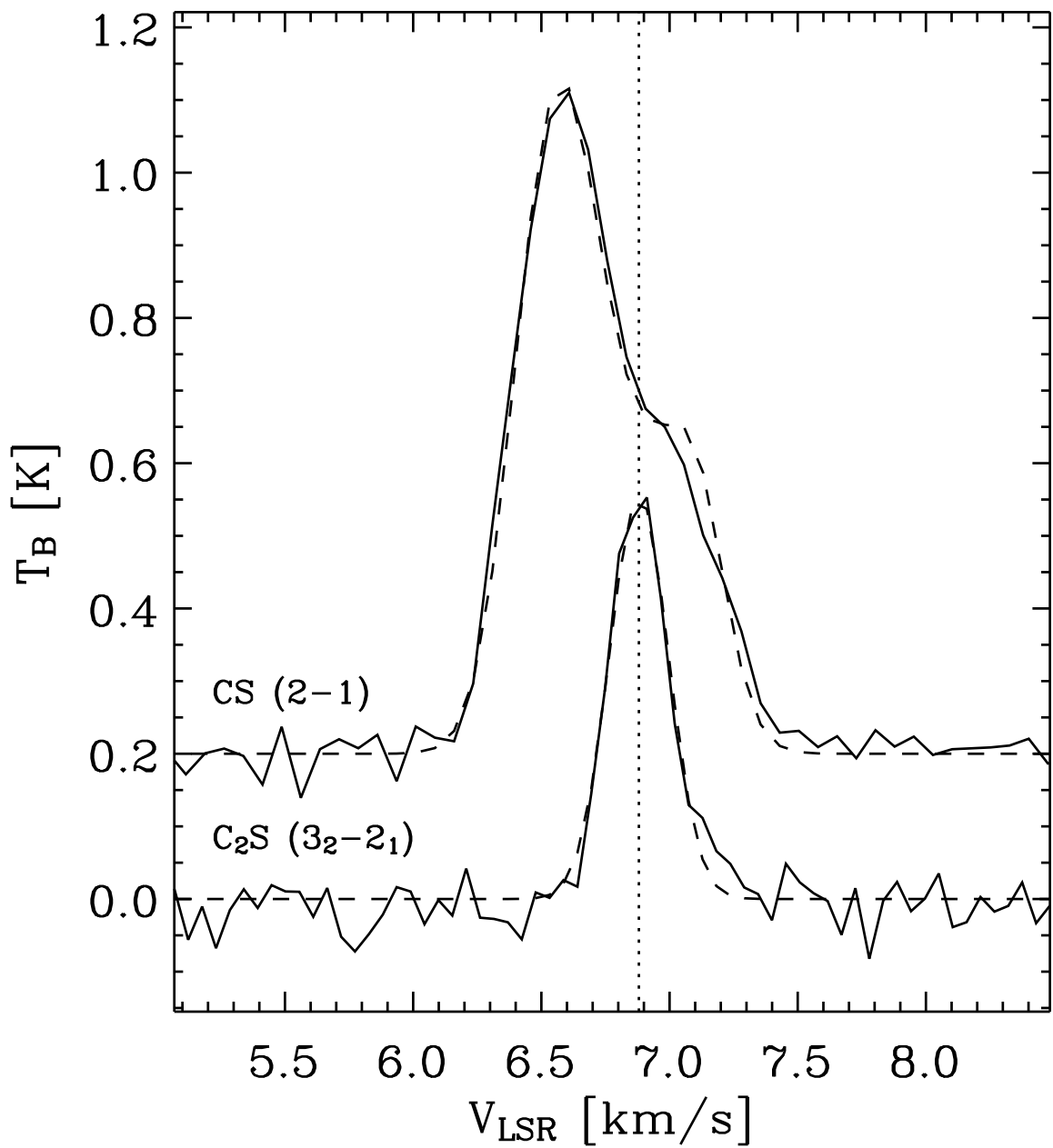}
\caption{CS and {\cts} spectra averaged over the half maximum contour
  of {\cts} emission in {\core}. The vertical dotted line shows the
  central velocity of the infalling layer determined from the {\hill}
  fit to the CS line profile and is seen to be nearly identical to the
  velocity centroid of the {\cts} line profile fit by a single
  Gaussian component. \label{ccs-cs}} 
\end{figure}

Figure~\ref{ccs-cs} shows composite spectra of the CS and {\cts} lines
averaged within a half maximum power contour of the {\cts} emission (see
Figure~\ref{velintcs}). The composite CS spectrum shows a blue peaked, 
asymmetric profile. The best fit to this spectrum using the {\hill}
model returns an infall velocity of $0.14$\,{\kms}, nearly identical
to the velocity centroid of the {\cts} emission.

This supports the idea that the {\cts} emission is tracing the
CS absorption layer, but it still remains unclear why there is no blue
component to the {\cts} profile. Chemistry is likely responsible.

\section{Conclusions} \label{sConc}
High spectral resolution maps of CS and {\nthp} taken with the FCRAO
14\,m telescope are fit with two distinctly different radiative
transfer models to gain insight into the dynamical and chemical
evolution of the recently discovered PPC in Taurus, {\core}.

The analytic and numerical models both suggest that {\core} is
in the early stages of dynamical evolution. Infall velocities of
$\gtrsim 0.1$\,{\kms} are needed in the outer regions of {\core} to
fit our data adequately. This translates to an accretion rate of $\sim
10^{-6}$\,{\msun}\,yr$^{-1}$, uncertain to within a factor of 5. The
observed dynamics are not due to gravitational
collapse and are not consistent with the standard picture
for low mass star formation. The extent and shape of the asymmetric CS
line profiles are more consistent with planar accretion.

There is modest evidence for chemical depletion near the highest
density regions of {\core} in CS and {\ceo}. This suggests that
{\core} is chemically young. Our models are not very sensitive to
chemical evolution and do not shed light on this issue.   

Fits to our CO isotopologue data yield a {\thco} to {\ceo} relative
abundance ratio of $\approx 10$. These fits also reveal the {\thco}
line centroid to be shifted relative to the other molecular species by 
7.1\,kHz or 0.02\,{\kms}. If this is due to an incorrect observing
frequency, the best estimate for the true transition frequency is
110.2013470\,GHz.

The {\cts} emission seen in Paper~I may be tracing the infalling CS
gas that is producing the observed self-absorption. This would explain
the apparent velocity offset of {\cts} with respect to the optically
thin, high density tracers. However, it is unclear why there is no
blue component to the {\cts} line profile.

The gas dynamics and distributions of molecular species in {\core} are
distinct from the complex processes related to the known young stellar
objects within the greater L1551 region and are evidence for continued
star formation in this evolved cloud.

\acknowledgments

Thanks to Mark Heyer for help with the data collection and
reduction. J. S. would also like to thank P. C. Myers for an
enlightening discussion and H. Dickel for the ``{\sc rt}'' code and 
correspondence throughout. We thank Al Glassgold for insightful
suggestions. This research was partially supported by
the National Science Foundation under grant AST 02-28963. 

Facilities: \facility{FCRAO 14~m Telescope}

\clearpage

\clearpage

\clearpage

\clearpage

\clearpage

\clearpage

\clearpage

\clearpage
\begin{deluxetable}{lc}
  \tablecolumns{2} 
  \tablewidth{0pt}
  \tablecaption{Overview of Numerical Cloud Models \label{da94table}} 
  \tablehead{
    \colhead{Model Parameter} & 
    \colhead{Value}  }
  \startdata 
  $n(r)$\dotfill  & $n_0/\left[1+\left(r/r_c\right)^\alpha\right]$ cm$^{-3}$ \\
  $n_{mol}(r)$\dotfill   & $X_{mol}(r)$\tablenotemark{\dagger}$n(r)$
  cm$^{-3}$ \\ 
  $v(r)$\dotfill  & $v_0\,(r/r_{min})^{p}$ {\kms}\\
  $\Delta v_{turb}$\dotfill   & 0.2\,{\kms} \\
  $T_K$\dotfill & 9\,K \\
  $r_{min}$\dotfill  & 67\,AU \\
  $r_{max}$\dotfill  & 0.2\,pc \\
  $r_c$\dotfill    & 0.05\,pc \\
  $\alpha$\dotfill    & 2.2 \\
  \enddata 
  \tablenotetext{\dagger}{$X_{mol}(r)$ is either constant with radius
  or takes the form of Equation~\ref{drop}.}
\end{deluxetable} 

\clearpage

\begin{deluxetable}{llll}
  \tablecolumns{4} 
  \tablewidth{0pt}
  \tablecaption{Family of Numerical Models \label{da94models}} 
  \tablehead{
    \colhead{Model Name} & 
    \colhead{Velocity Profile} & 
    \colhead{Abundance Profile} & 
    \colhead{Comments}  } 
  \startdata 
  \cutinhead{``Accretion'' Models}
  {\sc acv-i}   & $v(r) =$ constant  & $X_{mol}(r) =$ constant &
  Numerical analogue to {\hill} model. \\
  {\sc acv-ii}   & $v(r) =$ constant  & $X_{mol}(r) =$
  ``drop''\tablenotemark{a} &
  Like model {\sc acv-i} but with central \\*
  & & & depletion. \\
  {\sc a-i}   & $v(r) \propto r^{1/2}$  & $X_{mol}(r) =$ constant &
  High velocity gas at large core radii.\\
  {\sc a-ii}   & $v(r) \propto r^{1/2}$   & $X_{mol}(r) =$
  ``drop''\tablenotemark{a} & 
  Like model {\sc a-i} but with central \\*
  & & & depletion. \\
  \cutinhead{``Collapse'' Models}
  {\sc gc-i}  & $v(r) \propto r^{-1/2}$  & $X_{mol}(r) =$ constant & 
  Inside-out collapse model with \\*
  & & & constant abundance fraction. \\
  {\sc gc-ii}  & $v(r) \propto r^{-1/2}$  & $X_{mol}(r) =$
  ``drop''\tablenotemark{a} &  
  Inside-out collapse model with \\*
  & & & central depletion. \\
  \enddata
  \tablenotetext{a}{See Equation~\ref{drop}.}
\end{deluxetable} 

\clearpage
\begin{deluxetable}{lc}
  \tablecolumns{2} 
  \tablewidth{0pt}
  \tablecaption{{\hill} Model Fit to Average CS Spectrum
  \label{hillfittable}}  
 \tablehead{
    \colhead{Model Parameter} & 
    \colhead{Best Fit Value}  }
  \startdata 
  $\tau_C$\dotfill  & 2.02 \\
  $\sigma$\dotfill  & 0.15\,{\kms}\\
  $T_C$\dotfill     & 4.57\,K \\
  $v_{LSR}$\dotfill & 6.72\,{\kms} \\
  $v_{in}$\dotfill  & 0.15\,{\kms} \\
  RMSE\tablenotemark{a}\dotfill      & 2.03 \\
  \enddata 
  \tablenotetext{a}{The root mean square of the error as defined in
    Equation~\ref{rmse}.}
\end{deluxetable}

\clearpage
\begin{deluxetable}{lccccc}
  \tablecolumns{6} 
  \tablewidth{0pt}
  \tablecaption{Numerical Model Fits to CS Emission \label{daresults}} 
  \tablehead{
    \colhead{Name} & 
    \colhead{$n_0$} & 
    \colhead{$v_0$} &
    \colhead{$X_{CS}$} & 
    \colhead{$\tau_{max}$} &
    \colhead{RMSE\tablenotemark{a}} \\
    \colhead{} &
    \colhead{$10^4$\,cm$^{-3}$} &
    \colhead{\kms}  &
    \colhead{$10^{-9}$} &
    \colhead{} &
    \colhead{}  }
  \startdata 
  \cutinhead{``Accretion'' Models}
  {\sc acv-i}   & 4.6 & 0.16  & 0.8 & 2.8 & 0.95 \\
  {\sc acv-ii}  & 7.9 & 0.15  & 1.1 & 5.1 & 1.09 \\
  {\sc a-i} & 3.3 & 0.015 & 1.4 & 3.0 & 0.89 \\
  {\sc a-ii}  & 6.9 & 0.001 & 1.4 & 5.2 & 0.77 \\
  \cutinhead{``Collapse'' Models}
  {\sc gc-i}   & 4.2 & 0.5   & 1.1 & 5.1 & 2.38 \\
  {\sc gc-ii}  & 6.3 & 0.5   & 2.1 & 8.9 & 2.68 \\
  \enddata 
 \tablenotetext{a}{The root mean square of the error as defined in
    Equation~\ref{rmse}.}
\end{deluxetable} 

\clearpage
\begin{deluxetable}{lccccc} 
  \tablecolumns{6} 
  \tablewidth{0pt}
  \tablecaption{Numerical Model Fits to CO
  Isotopologue Emission \label{dacotable}}
  \tablehead{
	\multicolumn{6}{c}{\thco} \\
	\cline{1-6}
    \colhead{Name} & 
    \colhead{$X_{mol}$} & 
    \colhead{$\Delta v_{turb}$\,[{\kms}]} &
    \colhead{$T_k$\,[K]} &    
    \colhead{$\tau_{max}$} &    
    \colhead{RMSE\tablenotemark{a}} }
  \startdata 
  {\sc acv-i}   & $6.6(-7)$ & 0.39 & 8.8 & 3.1 & 4.19 \\
  {\sc acv-ii}  & $9.8(-7)$ & 0.39 & 8.7 & 4.9 & 4.11 \\
  {\sc a-i} & $8.6(-7)$ & 0.37 & 9.0 & 2.7 & 4.47 \\
  {\sc a-ii}  & $1.1(-6)$ & 0.38 & 8.8 & 4.7 & 4.18 \\
  {\sc gc-i}   & $7.9(-7)$ & 0.42 & 8.7 & 3.6 & 3.96 \\ 
  {\sc gc-ii}  & $1.3(-6)$ & 0.43 & 8.7 & 5.2 & 3.96 \\
	\cutinhead{\ceo} 			 
  {\sc acv-i}   & $6.9(-8)$ & 0.29 & 9.0 & 0.37 & 0.96 \\
  {\sc acv-ii}  & $1.0(-7)$ & 0.30 & 9.2 & 0.58 & 0.95 \\
  {\sc a-i} & $9.2(-8)$ & 0.24 & 10.3& 0.28 & 1.02 \\
  {\sc a-ii}  & $1.1(-7)$ & 0.28 & 12.0& 0.37 & 0.99 \\
  {\sc gc-i}   & $1.0(-7)$ & 0.34 & 6.1 & 0.97 & 0.94 \\ 
  {\sc gc-ii}  & $1.7(-7)$ & 0.35 & 6.1 & 1.4  & 0.94 \\
  \enddata 
  \tablenotetext{a}{The root mean square of the error as defined in
    Equation~\ref{rmse}.}
\end{deluxetable} 

\end{document}